# A Pareto Algorithm For Efficient De Novo Design Of Multi-Functional Molecules

Frits Daeyaert and Michael W. Deem
Departments of Bioengineering and Physics & Astronomy
Rice University
Houston, TX

## Abstract

We have introduced a Pareto sorting algorithm into Synopsis, a de novo design program that generates synthesizable molecules with desirable properties. We give a detailed description of the algorithm and illustrate its working in 2 different de novo design settings: the design of putative dual and selective FGFR and VEGFR inhibitors, and the successful design of organic structure determining agents (OSDAs) for the synthesis of zeolites. We show that the introduction of Pareto sorting not only enables the simultaneous optimization of multiple properties but also greatly improves the performance of the algorithm to generate molecules with hard-to-meet constraints. This in turn allows us to suggest approaches to address the problem of false positive hits in de novo structure based drug design by introducing structural and physicochemical constraints in the designed molecules, and by forcing essential interactions between these molecules and their target receptor.

## Introduction

De novo design of molecules is a technique to computationally design compounds with desirable properties. It combines a computational procedure or scoring function to predict these properties with a molecule generator and optimizer that produces molecules that score well in the scoring function. De novo design is most often applied in the field of drug design, but if the scoring function used is independent from the molecule generator it is generally applicable in materials design. This is the case for the de novo program Synopsis, which has been successfully applied to drug design (Tsimafeyeu et al., 2014), the design of organic structure determining agents (OSDAs) for zeolites (Pophale et al., 2013) and the design of linkers for metal organic frameworks (MOFS) (Bao et al., 2015).

An issue faced by any de novo design algorithm is the synthetic accessibility of the designed compounds. In the Synopsis program this is addressed in a very direct way by not only producing the structure of the designed molecules, but also generating a synthesis route by well-documented organic chemistry reactions from commercially available or proprietary starting materials (Vinkers et al., 2003).

The score that rates the desirability of a compound is difficult to express as a single number (Segall, 2014). In the original genetic algorithm used by Synopsis this has been addressed by using a step-wise scoring function. A number of properties are used as filters, and the final and computationally most expensive property is only calculated for those molecules that pass these filters. At the start of a Synopsis run, the genetic algorithm it uses as an optimizer searches for molecules that pass these filters one by one. After a certain number of function evaluations, the working population of the genetic algorithm contains molecules that pass all the filters, and progress is made in optimizing the final property.

However, when a larger number of more elaborate property filters are used in a design project, we have observed that convergence of this step-wise procedure is unsatisfactory. Additionally, in the stepwise scoring procedure, there is only one single final score that will eventually be optimized. We, and others (Besnard et al., 2012), have recently encountered design problems where more than one property needs to be optimized. Examples include the design of OSDAs for chiral zeolites, and the design of molecules that have either to bind to 2 or more biological targets, or have to bind specifically to one particular target and not to another.

We have therefore set out to adapt our structure optimization algorithm to address the multi-objective nature of the molecular design problem. Generally, two approaches have been used for this (Nicolaou et al., 2012). One is to condense the multiple scores into a single one using a weighing scheme. A priori weights are assigned to the properties, and the weighed sum of scores is optimized as a single objective function. The other approach is to use a Pareto optimization algorithm based upon the concept of non-dominance. The end result of such an algorithm is a family of compounds that are optimal in a Pareto sense, which means that there exist no compounds in that set that score better on all individual scores. Because the Pareto approach is very straightforward to introduce in the genetic algorithm used by Synopsis, and because it does not depend on a weighing scheme that has to be set up for each individual molecular design effort, we have opted for this latter approach.

We describe the details of the new algorithm in the methods section. To illustrate its working we present the results of two typical molecular design efforts, one in a drug design setting involving 2 different biological targets, and one in the field of organic structure directing agents (OSDAs) for the synthesis of zeolites.

# METHODS

In the methods section we will first describe the original Synopsis algorithm and then detail the recent adaptations to more efficiently handle multi-objective optimization. A summary description of the scoring functions that we have chosen to illustrate the working of Synopsis with will also be provided.

## SERIAL OPTIMIZATION

The search space of Synopsis is defined by the set of molecules that can be generated by applying a list of pre-defined well-documented organic chemistry reactions to a user-supplied database of commercially available or proprietary reagents. A genetic algorithm is used to find combinations of reactions and reagents that optimize the scoring function for the resulting products. The output of a Synopsis run is therefore a set of molecules that score well on the scoring function, together with their synthesis route by common reactions starting from available starting materials.

The set of reactions available to synopsis presently comprises 91 reactions. The set of reagents to be used in these reactions is organized as a 'shelf'. For each molecule the reactions in which it can participate are determined and stored on disk. To determine these reactions, both the presence of functional groups that are required for the reactions and the absence of interfering functional groups are checked on a per-reaction basis. A similar method to automatically generate molecules has been adopted by Hartenfeller et al. (Hartenfeller et al., 2012), but here the reagents are not screened on the absence of interfering functional groups.

The optimization engine of Synopsis is a genetic algorithm. A chromosome of the genetic algorithm consists of a tree data structure that has reagents and reaction products for nodes, connected by edges that represent reactions. The maximal depth of the tree defines the number of reactions that is allowed to form one solution molecule. To limit the synthetic complexity of the generated solutions this depth is typically set to between 3 and 5.

At the start of the algorithm, a population of random solutions is generated by randomly selecting reactions and reagents and tree depths. The population size is typically 100. When the population is complete it is evolved by applying one of 6 operators to one or more parent chromosomes to form a child chromosome:

'add': addition of a reaction step

'cut': deletion of a reaction step

'replace random': replacement of a reagent by another, randomly selected reagent

'replace like': replacement of a reagent by another, similar reagent

'combine': combination of 2 synthesis routes

'random': replacement of the chromosome by a completely new random chromosome

The parent chromosomes are selected using tournament selection: a set of N chromosomes is randomly selected, and the one with the best score is chosen as the parent. Operators 1 through 4 require one parent, operator 5 requires 2 parents, and operator 6 does not require a parent. In the 'replace random' operator, the new reagent is randomly selected from the shelf, with the restriction that it has to have the correct functionality to participate in the reactions defined by the chromosome. In the 'replace like' operator, the new reagent must additionally have a minimal similarity to the replaced reagent. The similarity is expressed by a Tanimoto coefficient based on a 1024 binary string of 2D descriptors, the similarity threshold used is typically .75. In the 'combine' operator, the 2 selected parents are searched for chemical functionalities that allow them to participate in a reaction. If no such functionality is detected, 2 other parents are selected. If more than one functionality is detected, one of the allowed reactions is selected at random.

Child chromosomes are produced and evaluated one by one. This 'steady state' mechanism as opposed to the more classical 'generational replacement' generally adopted in genetic algorithms allows for a very straightforward parallelization of the genetic algorithm. After evaluation of the child's score, it is compared to the score of the worst scoring chromosome in the population. If the child's score is better, it replaces the worst scoring chromosome, otherwise the child is discarded.

The scoring of a molecule is typically a multistep process where a number of easy-to-calculate properties are calculated first and used as a filter. If a molecule passes all the filters, the final property that is to be optimized is calculated. A scoring function to be used by synopsis must be a user-provided stand-alone program or script that reads a molecular data file from disk and that produces an output file containing the score of the molecule in the form of the numerical values of the list of calculated properties of the input molecule. For each property in this score vector a score type must be specified that determines how Synopsis is to treat the property value. For maximal flexibility, 8 score types are available as listed in table 1. Six score types define filters, and 2 score types define an optimization type, i.e. minimize or maximize. To determine the winning chromosome in the tournament selection, the properties are evaluated in a step-wise fashion. First, the filter-type scores are compared one by one in a fixed order to evaluate whether one molecule passes the filter and the other does not. If both molecules pass all the filters, the final property, which must be either of the minimize or of the maximize type, will decide the winner. The number of scores and the score type of each score are provided in a score definition file that is communicated to both the molecule generator and the scoring function program. The only requirements for the program or script that calculates the molecular scores is that it reads from disk the molecular data file and the score definition file, and that it writes to disk the score in the form of a simple list of numbers. This allows the use of any third-party or proprietary software to be used in combination with Synopsis.

| **Score Type** |
| --- |
| binary |
| greater than a threshold |
| less than a threshold |
| greater than or equal to a threshold |
| less than or equal to a threshold |
| bracketed between a minimum and a maximum value |
| minimize |
| maximize |

Table 1: Score types defined for Synopsis.

Communication between Synopsis and the user-provided scoring function program is managed by a separate scoredeamon program and relies on disk file semaphores. Multiple of these daemons can be started concurrently on different computers, allowing an efficient parallelization as the steady-state replacement mode of the genetic algorithm does not require synchronization at any point.

The output of a Synopsis run can be converted to an html document to be viewed with a web browser. It lists all generated compounds sorted according to their scores. By clicking onto each entry a link is provided to its synthesis route.

## PARETO OPTIMIZATION

To allow the simultaneous minimization or maximization of multiple molecular property descriptors we have replaced the serial comparison procedure used in the tournament selection by a Pareto non-dominance procedure. To do so, we have adopted the non-dominated sorting procedure described by Deb et al. (Deb et al., 2002). We have previously applied this algorithm to the alignment of flexible molecules (Daeyaert et al., 2005). With this algorithm, the population of the genetic algorithm is sorted in 2 phases. First, the different Pareto fronts are identified and used to sort the population according to their Pareto front number. This is done by determining the Pareto front for the entire population, and then finding the next Pareto front after the previous one is removed from the population. This procedure is repeated until all fronts have been determined. A Pareto front is defined as a set of solutions in which no solution is dominated by any other solution in the set. A solution is not-dominated by another solution if there is no score on which it scores better than the other. In a second phase, the solutions within each Pareto front are sorted using a 'crowding distance' metric. This is a measure of the distance in scores between a solution and the solutions of which the scores are closest. Its purpose is to evenly spread the solutions within the Pareto front. The crowding distance metric is straightforwardly defined for property values that have to be minimized or maximized (Deb et al., 2002). However, especially at the beginning of a Synopsis run, many solutions are present that do not pass some of the filters and for which the final property values that are to be minimized or maximized have not been calculated. For these molecules we use a different secondary sorting method based on the order in which they have been generated and entered the population. The rationale for this is that 'older' chromosomes have had their time to reproduce and that the 'younger' chromosomes have more potential to drive evolution and explore new areas of the search space.

## DUAL AND SELECTIVE FGFR/VEGFR INHIBITORS WITH AUTODOCK-VINA

To explore the performance of the Pareto optimization algorithm in a drug design setting we have set up a scoring function that calculates the binding of a molecule to the Fibroblast Growth Factor Receptor (FGFR) and the Vascular Endothelial Growth Factor Receptor (VEGFR). Both dual and selective inhibitors of these receptors are of interest as treatments for cancers in which they are up regulated (Renhowe, 2002). To predict the binding energy of molecules to available crystal structures of the receptors we have used the open-source program Autodock-Vina (Trott, 2011). All dockings are performed in the rigid receptors with fully flexible ligands. The exhaustiveness parameter is set to 32, and the search space is restricted to a cubic area of side 15 Å centered at the binding site. The receptor coordinates have been extracted and prepared for docking with Autodock-Vina using the appropriate python script from AutoDock MGLTools toolkit (Morris, 2009). To select a crystal structure for docking putative VEGFR inhibitors we have used the IChem IFP program that calculates interaction fingerprints for a ligand bound to a receptor (Marcou, 2007).

In addition to the binding energies to both receptors, a number of additional scores are calculated to be used as filters. To limit the flexibility of the molecules both the total number of rotatable bonds and the length of the longest chain of sp3-sp3 linkages in a molecule are determined and upper bounds are set for both properties. To enforce the absence of unwanted reactive chemical groups in the generated molecules a binary flag is set to indicate their presence. As unwanted reactive groups we define acid chlorides, acid anhydrides and sulfonic chlorides.

The molecular volume is approximated as an integral over atom-centered Gaussians as in (Grant, 1996). A lower bound to the molecular volume is set to avoid attempts to docking molecules that will be too small to efficiently fill the binding site. A high-volume bound is set to prevent the emergence of very large molecules that would get large docking scores through interactions with the protein away from the binding site.

LogP, the polar surface area and the molar refractivity are obtained using the Open Babel program obprop. To improve the drug-likeness of the generated molecules we have adapted the ranges proposed by (Ghose, 1999) for these properties as filters.

To enforce specific interactions of the designed ligands with individual residues of the receptors we have used an approach adopted from the concept of interaction fingerprints (Deng, 2004, Marcou, 2007). Using the geometric rules described in (Marcou, 2007), for each residue in the binding pocket the presence of a hydrophobic interaction, a H-bond donor interaction, or a H-bond acceptor interaction of the docked ligand with either the backbone or the side chain of the residue is detected and stored as a bit. We then compare the sums of a number of selected subsets of these interaction bits with the corresponding sums obtained from the PDB structure of the complex with the original active ligands. Thus, we can enforce that in the complex of a de novo designed molecule a minimal number of similar interactions are present as in the known active compounds.

### ORGANIC STRUCTURE DIRECTING AGENTS (OSDAS) FOR ZEOLITES

We have previously reported the use of Synopsis to generate ODSAs for Zeolites (Pophale, 2013). The final score to be optimized here is the non-bond Lennard-Jones interaction energy of the OSDA with the zeolite and with other OSDASs in the zeolite. This is obtained by first fitting a number of copies of the OSDA into the zeolite using a fast Fourier procedure, followed by a molecular dynamics simulation. Full details on the computational procedure have been given in (Pophale, 2013). As additional structural properties we determine the number of rotatable bonds, the length of the longest chain of sp3-sp3 linkages, the molecular volume, the absence of other atoms than C, N or H, the presence of absence of unwanted chemical functionalities and the formal charge of the molecule.

# RESULTS

### FGFR/VEGFR INHIBITORS

In setting up a dual FGFR/VEGFR scoring function for synopsis we have set out from the crystal structure of the dual FGFR/VEGFR inhibitor E3810 (Lucitanib) bound to the ATP binding site of FGFR. The pdb entry is 4RWL (Sohl, 2015). No crystal structure of a dual FGFR/VEGFR inhibitor bound to VEGFR was available at the time of inception of the present study. Therefore we have downloaded 13 high-resolution crystal structures of VEGFR-inhibitor complexes from the PDB and selected the one that best accommodates E3810 (see supplementary information).

The binding of E3810 to FGFR is mediated by a number of H-bond and hydrophobic interactions. The ligand acts as an H-bond acceptor for the backbone N-H of Ala 564 in the hinge region of the receptor and for the backbone N-H Asp 641 in the backpocket of the receptor. Using the IChem IFP program (Marcou, 2007) we have identified the hydrophobic interactions between the ligand and the receptor. In order to obtain a more fine-grained description of the interactions we have written a program to separate these into side chain and backbone interactions. Thus we have identified 25 hydrophobic interactions between E3810 and FGFR. To enforce proper interaction between the de novo generated molecules and FGFR we have set the requirement that the H-bond interaction with the hinge region, the H-bond interaction with the back pocket, and at least 18 (2/3) of the 25 hydrophobic interactions must be present for a molecule to be considered active. In a number of known FGFR inhibitors the H-bond interaction with the hinge region is present as an H-bond donor interaction towards the carbonyl backbone of either Ala564 or Glu562 in addition to or instead of the acceptor interaction with N-H Ala564. We therefore have allowed the hinge H-bond interaction to be alternatively present as one or more of these interactions.

In a similar way we have analyzed the interactions between the 2XIR VEGFR receptor structure and the docked E3810 and set up constraints for the design of VEGFR inhibitors. For VEGFR binding we require at least one of three possible H-bond interactions with the hinge region, an H-bond donor interaction with Asp 1046, and at least 14 out of a set of 22 hydrophobic interactions with the receptor.

The resulting score vector that is generated by the scoring function for the design of dual FGFR/VEGFR inhibitors is summarized in table 2. When a molecule is generated by synopsis, the total number of rotatable bonds, the length of the longest chain of sp3-sp3 linkages, and the presence of reactive groups are obtained from the 1D molecular structure and checked against the criteria. When the molecule passes these filters, the 3D coordinates of the lowest-energy conformation are generated using an ant algorithm (Daeyaert, 2007). From the 3D coordinates the molecular volume, logP, polar surface area and molar refractivity are calculated and compared to their threshold values. When the molecule passes these additional filters it is docked into the FGFR and VEFGR structures with Autodock-Vina. From the docked structures, the numbers of hinge and backpocket H-bond interactions and overall hydrophobic interactions for both receptors are determined and attached to the score vector. When the required numbers of interactions are not present for one of the 2 predicted binding modes, the corresponding score is scaled down by 10. The thus obtained scores form the last entries of the score vector that is past to Synopsis and used to Pareto-sort the population. Good scoring dual inhibitors will have the required structural and physicochemical properties, make H-bond interactions with the hinge and backpocket of both FGFR and VEGFR, make a minimum number of hydrophobic interactions with both receptors, and have a good Autodock-Vina score on both FGFR and VEGFR.

| score description | score type | threshold 1 | threshold 2 |
|---|---|---|---|
| rotatable bonds | l.e. threshold | 4 | |
| longest sp3-sp3 chain | l.e. threshold | 4 | |
| reactive groups | binary | | |
| volume | bracketed | 150. $Å^3$ | 400. $Å^3$ |
| logP | bracketed | -0.4 | 5.6 |
| polar surface area | l.e. threshold | 140. $Å^2$ | |

| molar refractivity | bracketed | 40. cm³/mol | 130. cm³/mol |
|---|---|---|---|
| hinge interaction FGFR | g.e. threshold | 1 | |
| backpocket interaction FGFR | g.e. threshold | 1 | |
| hydrophobic interaction FGFR | g.e. threshold | 18 | |
| hinge interaction VEGFR | g.e. threshold | 1 | |
| backpocket interaction VEGFR | g.e. threshold | 1 | |
| hydrophobic interaction VEGFR | g.e. threshold | 14 | |
| Autodock-Vina score FGFR | minimize | | |
| Autodock-Vina score VEGFR | minimize | | |

Table 2: Score vector used for the design of dual FGFR/VEGFR inhibitors.

The results of performing 3 Synopsis runs to design dual FGFR/VEGFR inhibitors using 30000 scoring function evaluations are summarized in the upper part of table 3. The molecules are named according to the order in which they have been generated. In all 3 runs, molecules with a predicted binding energy below -9.5. kcal/mol have been generated. Figure 1 shows a scatterplot of the predicted binding energies of the molecules in the final population of one Synopsis run. The first 3 Pareto fronts are shown as solid lines. There is considerable variation in the outcome and the course of the 3 runs. Figure 2 illustrates the Pareto-optimal fronts present in the final population of the 3 runs. As can be seen in table 3 the first molecules that comply with all structural and physicochemical constraints and that interact properly with both receptors appear after 4000 to 10000 function evaluations. The newest Pareto-optimal molecules join the Pareto front as early as after 11000 function evaluations and as late as after almost 30000 function evaluations. The 2D structures of the molecules that form the Pareto-optimal fronts of the 3 runs are shown in table 4. As an example, Figure 3 shows the proposed synthesis route of Syn026297. The synthesis routes for the other Pareto-optimal molecules are provided as supplementary information.

| Run | Pareto/Best | First | Last/Best |
|---|---|---|---|
| DUAL 1 | 4 | Syn004227 | Syn029725 |
| DUAL 2 | 3 | Syn004441 | Syn022114 |
| DUAL 3 | 1 | Syn009801 | Syn011303 |
| FGFR 1 | -11.2 | Syn006978 | Syn020386 |
| FGFR 2 | -10.9 | Syn007654 | Syn022753 |
| FGFR 3 | -12.6 | Syn003808 | Syn017721 |
| VEGFR 1 | -12.1 | Syn001887 | Syn021106 |
| VEGFR 2 | -12.6 | Syn000383 | Syn026071 |
| VEGFR 3 | -11.7 | Syn001406 | Syn022800 |

Table 3: Summary of Synopsis runs for dual and selective FGFR/VEGFR inhibitors. Column 2 presents the number of Pareto optimal solutions for the dual and the best score found for the selective inhibitors. Column 4 presents the last compound that has appeared in the dual inhibitor Pareto optimal front or the best selective inhibitor.

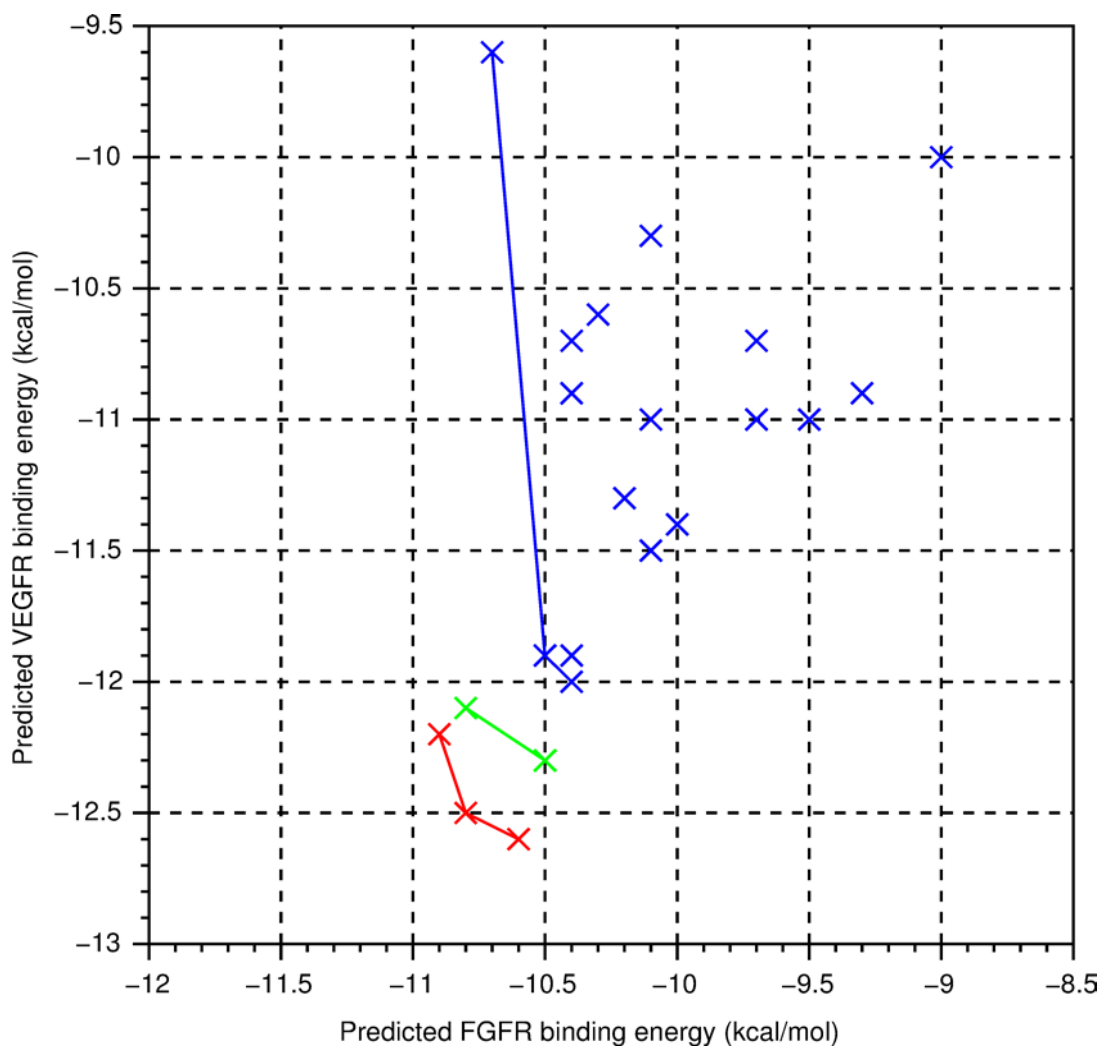

Figure 1: Scatter plot and Pareto fronts of the predicted FGFR and VEGFR binding energies of the molecules in the final generation of a selected Synopsis run. The lines connect the solutions comprising the first 3 Pareto fronts.

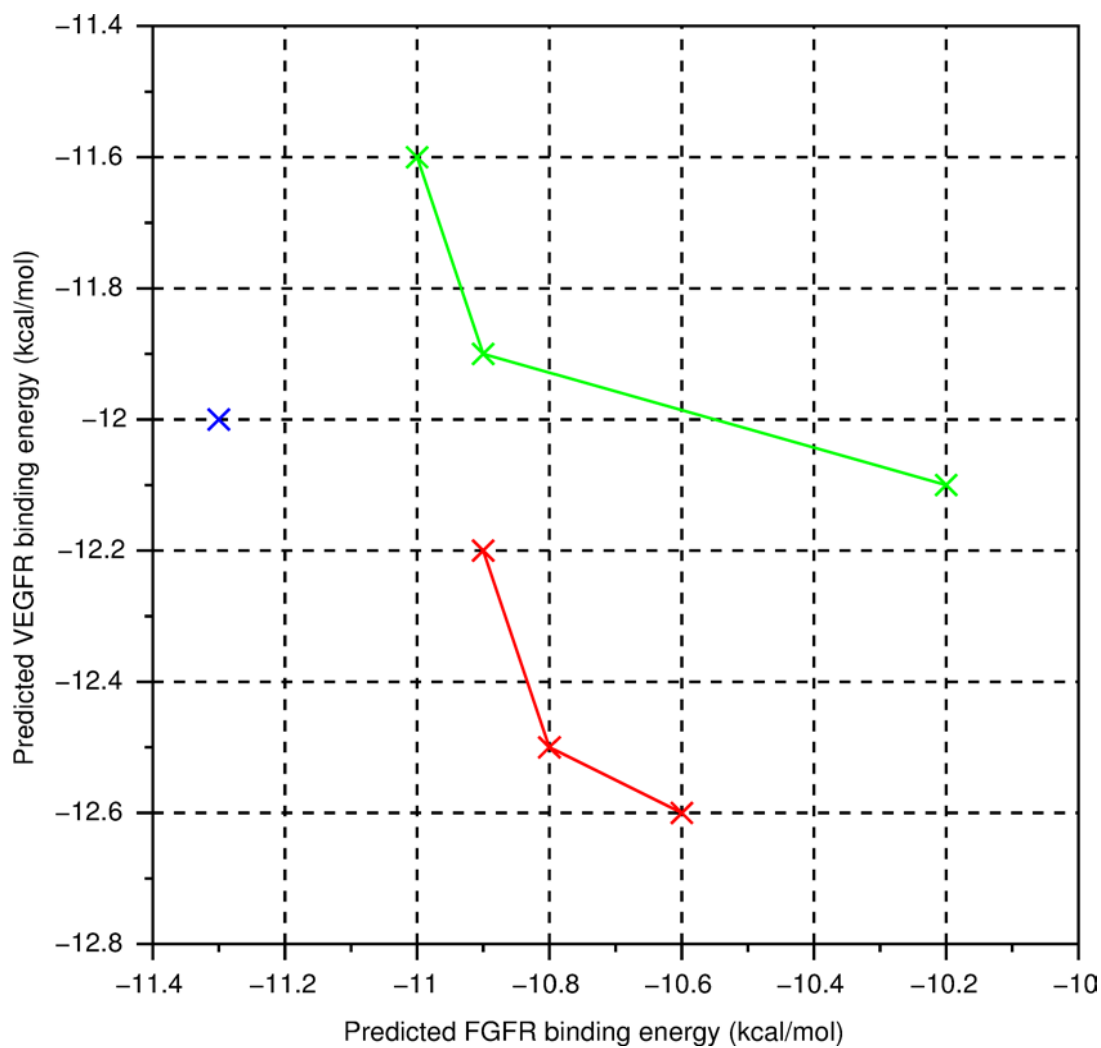

Figure 2: Pareto fronts of dual FGFR/VEGFR inhibitors generated in 3 separate Synopsis runs.

| Run | Molecule | FGFR | VEGFR | Structure |
|---|---|---|---|---|
| 1 | Syn026297 | -10.6 | -12.6 | |
| 1 | Syn026170 | -10.9 | -12.2 | |
| 1 | Syn027138 | -10.8 | -12.6 | |

| 2 | Syn006169 | -10.2 | -12.1 | 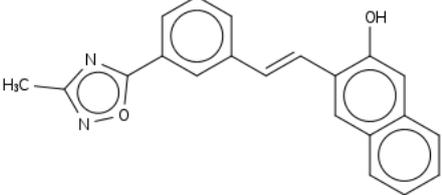 |
|---|---|---|---|---|
| 2 | Syn017054 | -10.9 | -11.9 | 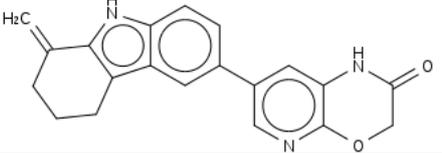 |
| 2 | Syn022114 | -11.0 | -11.6 | 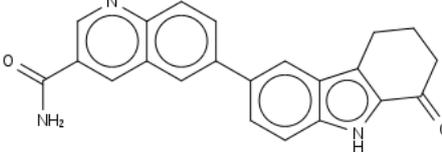 |
| 3 | Syn011303 | -11.3 | -12.0 | 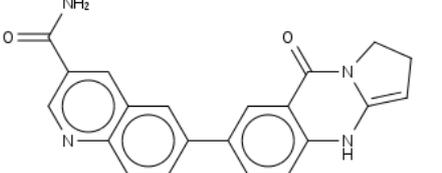 |

Table 4: Pareto-optimal putative dual FGFR/VEGFR inhibitors that have been generated in 3 Synopsis runs. Columns 3 and 4 contain the binding energies in kcal/mol as predicted by Autodock-Vina.

# Synthesis of Syn006169

Syn006169: 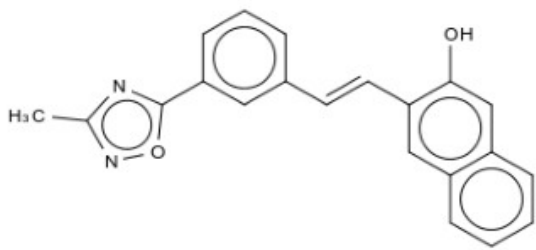

reaction = ALKYNETOALKENE
reagent(s):

Int011632 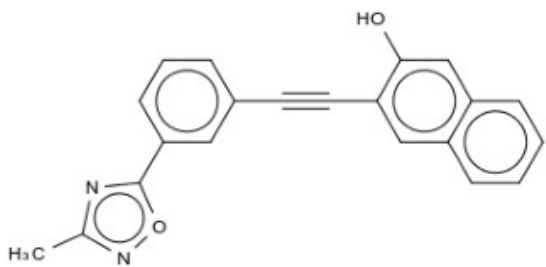

---

Int011632: 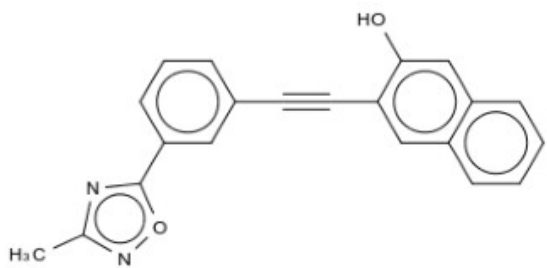

reaction =
SONOGASHIRA
reagent(s):

Int010967 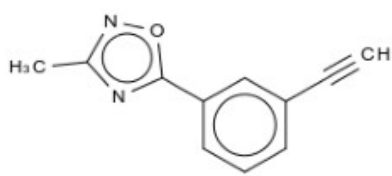    MFCD10000951 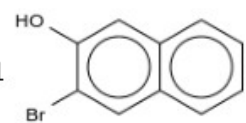

---

Int010967: 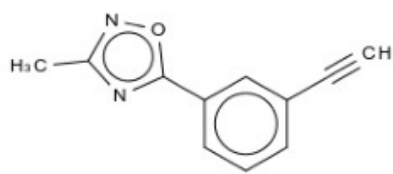

reaction = COREYFUCHS
reagent(s):

MFCD09817468 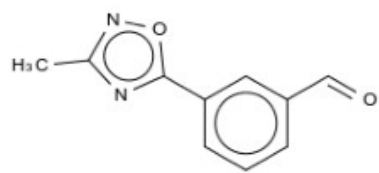

---

Figure 3: Synthesis route of Syn006169 as output by Synopsis.

It is found that in the 3 Synopsis runs the genetic algorithm has converged towards a population of compounds belonging to one or a few chemical families, each generated by a similar synthesis route and binding to the receptors in a similar way. In the first run, Syn026170 is generated in one step by an amide formation reaction from an amine and an acid. Syn026297 and Syn027138 are one-step modifications of this molecule. All three molecules bind to the hinge and backpocket regions of the receptors by the N-H of the acridinone moiety and the cyan group of the benzoxazole, respectively. This is illustrated for Syn026297 in Figure 4. In the second Synopsis run, Syn006169 is generated by the reduction of a Sonogashira adduct for which the required alkyne is provided by a Corey-Fuchs reaction. The naphtyl hydroxy group inteacts with the hinge region and the oxadiazole accepts the Asp N-H bond in the backpocket of the receptors. Syn022114 and Syn017054 are generated by a Suzuki-Myaura reaction leading to a substitued tetrahydrocarbazole that is further modified. The tetrahydrocarbazole NH provides the required H-bond to the hinge regions and the carbonyl groups at the other end of the molecule accept the backbone N-H in the backpockets. Syn011303 is the single Pareto-optimal compound in the 3th Synopsis run and is generated using a Suzuki-Myaura reaction followed by an amination reaction to form the amide substituent on the quinoline moiety. The NH on the quinazolinone provides the H-bond to the hinge region and the amide substituent on the quinolone accepts the backbone N-H in the backpocket of the receptors. Interestingly, the hydroxyquinoline part of Syn011303 of run 3 is identical to that of Syn022114. The non Pareto-optimal molecules in the final population of the genetic algorithm runs are for the most part close analogs to the ones in the Pareto front. The cross section between the sets of molecules in the 3 final populations is empty. In other words, in separate runs the algorithm has explored different regions of the search space.

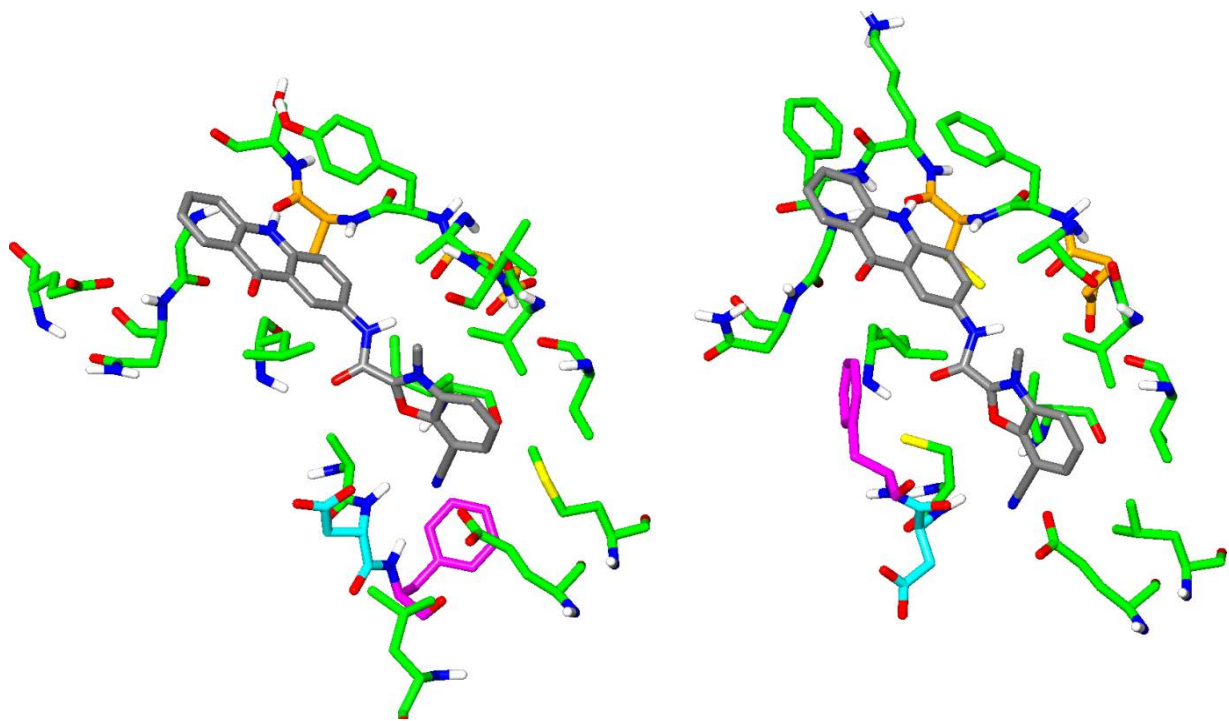

Figure 4: Predicted binding modes of Syn026297 to FGFR (left) and VEGFR (right). The carbon atoms of the residues in the hinge region of the receptor are colored orange, the carbon atoms of the Asp in the back pocket are in cyan, the carbon atoms of Phe in the DGF motif are in magenta.

To design selective inhibitors for FGFR vs. VEGFR we have adapted the scoring function to be used by Synopsis so that the FGFR binding energy is to be minimized and the VEGFR binding energy is to be maximized. As for the design of dual inhibitors, the FGFR score is scaled down if one of the 3 interaction criteria has not been met. The VEGFR score is now scaled down only when neither of the interaction criteria has been met. Good scoring selective FGFR vs. VEGFR inhibitors will thus have the required structural and physicochemical properties, make a H-bond interaction with the hinge and the backpocket of FGFR, make a minimum number of hydrophobic interactions with FGFR, be unable to make a H-bond interaction with the hinge or backpocket of VEGFR, make less than a maximum number of hydrophobic interactions with VEGFR, and have a good Autodock-Vina score on FGFR. For the design of selective VEGFR vs. FGFR inhibitors the scoring function has been adapted in an analogous way.

As shown in table 3, Synopsis is able to generate high scoring putative selective FGFR and VEGFR inhibitors. As with the dual inhibitors, different runs generate molecules belonging to 1 or a few chemical families, and there is no overlap between the sets of compounds generated in the runs. Table 5 shows the 2D structures of the most active selective compounds generated in each of the 3 runs. Figure 5 shows the 3D structures with the binding mode of the best scoring putative FGFR and VEGFR inhibitors found, Syn017721 and Syn026071, respectively.

Syn017721 provides a double H-bond interaction with the hinge region of FGFR through its amide functional group that is locked in the cis configuration in the 7-membered ring. The H-bond with the backpocket is provided by the cyan group. Syn026071 accepts H-bonds from the hinge and backpocket through its keto linker and amide group. The specificity of Syn017721 for FGFR stems from the 'DFG in 'conformation of the receptor structure of 4RWL (Sohl, 2015). The 2XIR receptor structure, which we have chosen as a model because it best accommodates the known dual receptor E3810 (cf. supplementary information), is in the 'DFG-out' conformation. Syn026071 and the other predicted selective FGFR inhibitors cannot bind the VEGFR receptor in this conformation as they would take in the position of Phe 1046 (Figure 5). The selectivity of the predicted VEGFR inhibitors can be attributed to their tight fitting in the backpocket of the receptor. When docked into the FGFR structure they will not effectively occupy the binding site and fail to make the correct interactions with the hinge and backpocket.

The synthetic routes generated by Synopsis for the other selective inhibitors are supplied as supplementary information.

| Run | Compound | Binding Energy | 2D structure |
|---|---|---|---|
| FGFR 1 | Syn020386 | -11.2 | |
| FGFR 2 | Syn022753 | -10.9 | 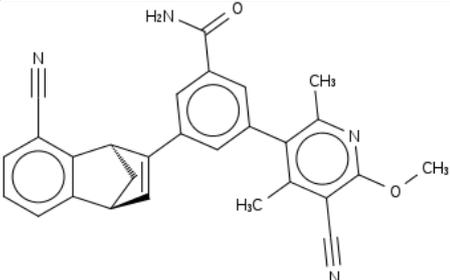 |
| FGFR 3 | Syn017721 | -12.6 | 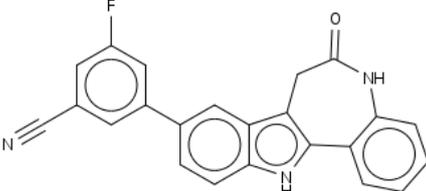 |
| VEGFR 1 | Syn021106 | -12.1 | 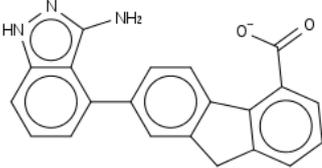 |
| VEGFR 2 | Syn026071 | -12.6 | 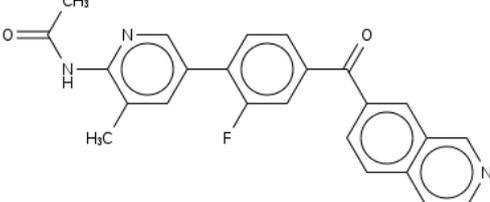 |
| VEGFR 3 | Syn022800 | -11.7 | 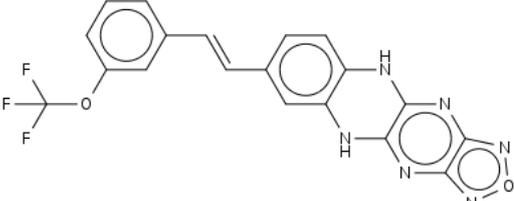 |

Table 5: Putative selective FGFR and VEGFR inhibitors generated in 2x3 Synopsis runs.

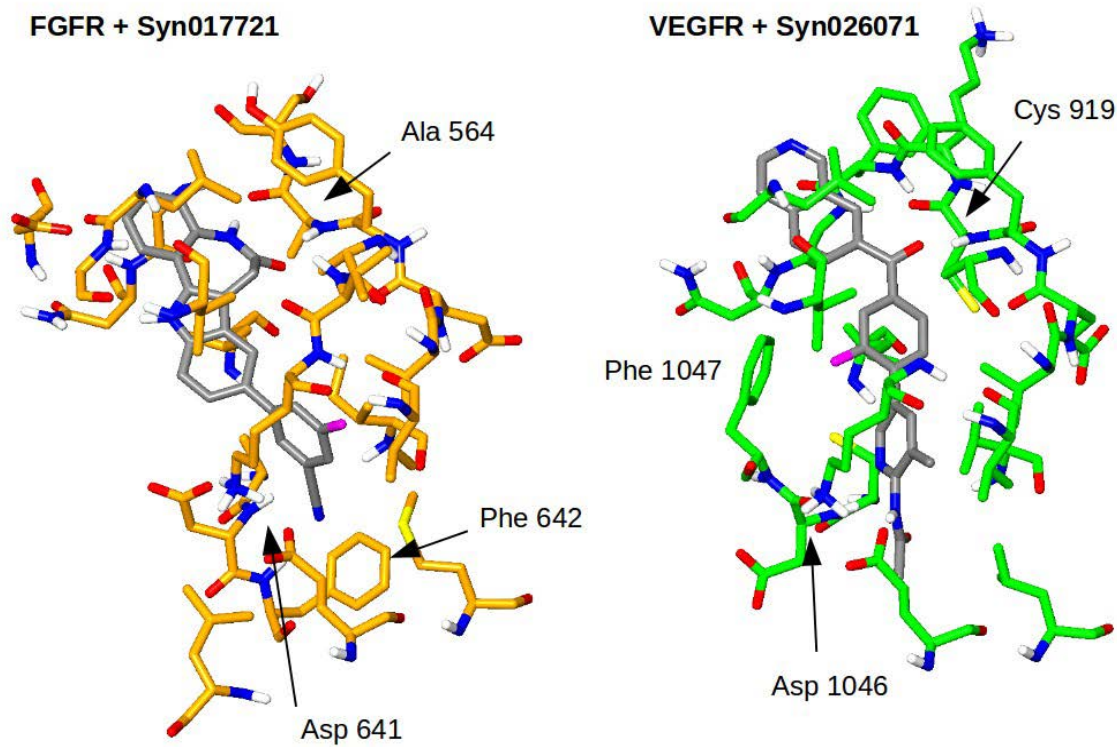

Figure 5: Predicted binding modes of the best scoring putative selective FGFR and VEGFR inhibitors.

## OSDAs

The score vector that is generated by the scoring function for the design of ODSAs for zeolites is summarized in table 6. Initially, the number of rotatable bonds, the molecular volume, the presence of atoms other than C, N or H, the presence of triply bonded C and the formal charge of a molecule generated by Synopsis are determined. If all of these properties are within their specific range, a number of copies of the OSDA is fitted into the zeolite. If this is successful, the molecular dynamics procedure is called to predict the stabilization energy which is the score to be minimized.

| score description | score type | threshold |
| --- | --- | --- |
| rotatable bonds | l.e. threshold | 1 |
| volume | l.e. threshold | 800 Å$^3$ |
| non-C, -N or –H atoms | l.e. threshold | 0 |
| triply bond C | l.e. threshold | 0 |
| total charge | g.t. threshold | 1. |
| fast Fourier fit | binary | |
| stabilization energy | minimize | |

Table 6: Score vector used for the design of zeolite OSDAs

Table 7 shows the 5 top scoring putative OSDAs obtained in 3 runs in which 30000 compounds have been generated and fitted into the structure of zeolite SSZ-39 (AEI). They are all generated by either a Menshukin reaction between a tertiary amine and an alkyl halogenide, or by alkylating a secondary or tertiary amine with either methyl- or ethyl Iodide. The individual synthesis routes of the compounds in table 7 are presented as supplementary information. Syn009753 has been synthesized and used to successfully produce the AEI, as described elsewhere (Schmidt, 2015). In contrast to the FGFR/VEGFR runs, there is considerable overlap between the sets of molecules generated by different runs. The best scoring molecule, Syn001853 in table 8, has been generated in 2 of the 3 runs. Also, the best scoring molecules appear in an earlier stage of the runs. This can be attributed to the fact that with the present set of reactions the only way to generate positively charged molecules is either by the Menshutkin reaction or by the methylation or ethylation of amines.

| Molecule | Binding Energy (kcal/mol) | 2D structure |
|---|---|---|
| Syn001853 | -17.6 | 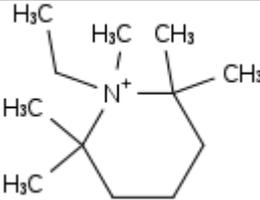 |
| Syn000211 | -17.5 | 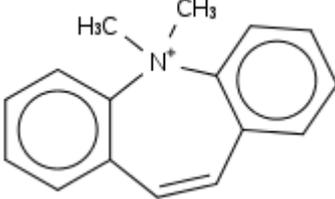 |
| Syn068142 | -16.9 | 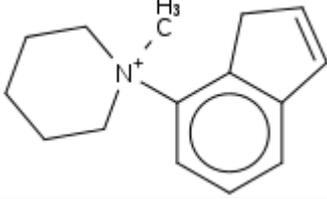 |
| Syn010248 | -16.7 | 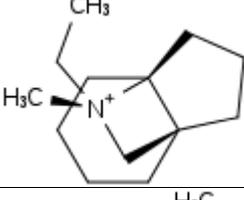 |
| Syn002582 | -16.7 | 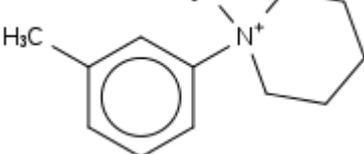 |

Table 7: 5 top scoring compounds in 3 separate runs using the OSDA scoring function.

## DISCUSSION

In the past, we have successfully deployed Synopsis in its serial optimization mode for drug design (Vinkers, 2003, Tsimafeyeu, 2014) and the design of an OSDA for zeolite HPM-1 (STW) (Schmidt, 2014). During the latter study it has emerged that it would be desirable to impose additional restrictions upon the structural properties of the designed compounds. More specifically, to ensure that the putative OSDAs are soluble and stable under the conditions required to produce commercially interesting zeolite types, they should contain only C, N and H, should not contain triply bonded C and must have a formal charge of at least +1. With these very specific restrictions added to the score function it has turned out that the serial optimization mode of Synopsis does not or only very sparingly produces good scoring compounds. We have therefore set out to replace the serial optimization by a Pareto optimization procedure. As we have shown in the results section this adaptation enables Synopsis to effectively generate good scoring compounds that obey the imposed restrictions. The success of the Pareto approach can be attributed to its inherent parallelism. In the serial optimization mode the algorithm has to address the different filters one by one. An improvement in the score vector at a position further than the first score that does not pass its threshold value will not be rewarded by the genetic algorithm. In contrast, in the Pareto mode any improvement in the score vector will lead to a fitter individual that can enter the population of the genetic algorithm.

An additional benefit of the Pareto approach is that it allows the simultaneous optimization of 2 or more properties without having to a priori define a weighing scheme. We have illustrated this by the design of putative dual and selective FGFR/VEGFR inhibitors. The resulting compounds have favorable predicted binding energies with their targets and additionally have structural and physicochemical parameters that are in ranges that are typical for drug molecules.

A possible issue with the Pareto approach is that when the number of dimensions of the multi-objective score becomes larger (>4), the number of Pareto-optimal solutions becomes intractable (Besnard et al. 2012). This problem is however alleviated in our scoring functions by defining many of the objectives as filters. The final Pareto optimal front contains molecules that pass all the filters for which the corresponding dimensions have condensed, and only those dimensions remain that correspond to the properties that are to be minimized or maximized.

The operators used to evolve the genetic algorithm that drives Synopsis are applied with equal probabilities. This appears to provide an efficient mix of optimization and exploration. Turning off either one of the operators completely deteriorates the performance of the algorithm, but otherwise it is quite insensitive to changes in the operator probabilities.

The scoring function we have used to generate putative OSDAs is well calibrated and has led to the successful synthesis of one of the generated compounds and its application to the synthesis of its target zeolite (Schmidt, 2015). In contrast, we have not calibrated or validated the scoring function for the FGFR/VEGFR inhibitors as this would be beyond the scope of the present methodological report. Nevertheless we like to highlight some of its features that can be useful in future real-life de novo drug design efforts. The AutoDock- Vina program is generally acknowledged to be a useful tool to predict the binding sides of active inhibitors and to enrich databases in virtual screening experiments (Biesiada, 2011). In virtual screening it is important to limit both false positive and false negative hits. False positive hits lead to useless in vitro screening of inactive compounds, while false negative hits may lead to the loss of promising drug candidates. We argue that in de novo design the false negative problem is less severe because of the vastness of the chemical space that is available. Using 1 reaction step in combination with a shelf comprising 52642 reagents, 154536 zero-order and 231540708 first-order reaction products can be formed by Synopsis. Thus, assuming a constant expansion factor, the number of compounds after N steps can be roughly estimated as $52642 \times [(231540708+154536)/52642]^N$. For N=5 steps, this is $\sim 10^{22}$ possible compounds, an impossibly large space to search combinatorially. Even when some potentially useful chemical classes of compounds are missed by searching this space, alternative hits will surely be discovered. The problem of false positive hits however is larger for a de novo design approach because the compounds have to be synthesized before testing. To alleviate the false positive problem we have added 2 types of filters to the de novo designed compounds before assigning the final Autodock-Vina scores. A first type of filter consists of a number of structural and physicochemical parameters. We have found that the use of these parameters not only improves the drug-like properties of the molecules, but also prevents Autodock-Vina from giving unrealistically high scores to either large hydrophobic molecules that make interactions outside of the actual binding site, or small and very polar molecules that make multiple hydrogen bonds to the receptor. With these structural and physicochemical filters in place, Autodock-Vina still assigns favorable binding energies to molecules that do not form interactions that are characteristic and deemed essential for known active inhibitors. We therefore have introduced additional constraints enforcing these interactions, for which we have used the geometrical criteria that are used to determine ligand-receptor interaction fingerprints (Marcou, 2007). As shown in the results section, the Pareto algorithm is able to produce molecules that fulfill these constraints.

## CONCLUSION

We have introduced a Pareto sorting algorithm into Synopsis, a de novo design program that generates synthesizable molecules with desirable properties. We have shown that this not only makes it possible to optimize multiple molecular properties simultaneously, but also greatly improves the ability to generate molecules with hard-to-meet constraints. We have also suggested approaches to alleviate the issue of false positive hits when using AutoDock-Vina, and by extension other docking programs, in a de novo design setting.

*SUPPLEMENTARY INFORMATION*

# SELECTION OF CRYSTAL STRUCTURES FOR THE MODELING OF FGFR/VEGFR INHIBITORS

In setting up a dual FGFR/VEGFR scoring function for synopsis we have set out from the crystal structure of the dual FGFR/VEGFR inhibitor E3810 (Lucitanib) bound to the ATP binding site of FGFR. The pdb entry is 4RWL (Sohl, 2015). From this structure the receptor coordinates have been extracted and prepared for docking with Autodock-Vina using the appropriate python script from AutoDock MGLTools toolkit (Morris, 2009). No crystal structure of a dual FGFR/VEGFR inhibitor bound to VEGFR was available at the time of inception of the present study. Therefore, we have downloaded 13 high-resolution crystal structures of VEGFR-inhibitor complexes from the PDB and used these to dock E3810 with Autodock-Vina. All dockings have been performed with fully flexible ligands and a fixed receptor. To select a suitable candidate receptor structure for predicting VEGFR binding from the thus obtained predicted VEGFR-E3810 complex structures we have used 3 criteria : the Autodock-Vina score, visual inspection and interaction fingerprints (tabel S1). The Autodock-Vina score we have used is the lowest predicted binding energy obtained using Autodock-Vina. The predicted complex structures were visually inspected and subjectively given a + or - score, based upon the orientation of the ligand in the receptors and on the presence of the appropriate receptor-ligand interactions. Finally, to obtain a quantitative measure of the predicted VEGFR-E3810 interactions we have used the IChem IFP script of (Marcou, 2007) to calculate the fingerprint Tanimoto similarity coefficient between the predicted complexes and the individual complex structures from the PDB. As seen in table S1, the docking of E3810 in the receptor structure of PDB entry 2XIR scores best on the 3 criteria and therefore of this structure was used for the prediction of VEGFR binding.

| PDB entry | Vina Score | Visual Inspection | IChem IFP |
|---|---|---|---|
| 2XIR | -11.5 | + | .75 |
| 3VHE | -11.3 | + | .50 |
| 4AG8 | -10.9 | + | .65 |
| 2RL5 | -10.7 | + | .53 |
| 3U6J | -10.7 | + | .45 |
| 3BE2 | -9.8 | - | .25 |
| 3CP9 | -9.8 | - | .16 |
| 3VHK | -9.8 | + | .31 |
| 3VID | -9.4 | + | .33 |
| 3B8Q | -9.0 | - | .36 |
| 3C7Q | -8.6 | - | .42 |
| 3DTW | -8.0 | - | .00 |
| 3EFL | -7.8 | - | .18 |

Table S8: Selection of a VEGFR receptor structure for the design of dual FGFR/VEGFR inhibitors.